\documentclass[epj]{svjour}
\usepackage{bm}
\usepackage{graphicx}

\begin{document}
\title{Anisotropy of the upper critical field in MgB$_2$: the two-gap Ginzburg-Landau theory}
\author{V. H. Dao and  M. E. Zhitomirsky}
\institute{Commissariat \`a l'Energie Atomique, DSM/DRFMC/SPSMS, 
17 avenue des Martyrs, 38054 Grenoble Cedex 9, France}
\date{Received: August 26, 2004/ Revised version: }

\authorrunning{V. H. Dao, M. E. Zhitomirsky}
\titlerunning{Anisotropy of the upper critical field in MgB$_2$}

\abstract{
The upper critical field in MgB$_2$ is investigated in the framework 
of the two-gap 
Ginzburg-Landau theory. A variational solution of linearized
Ginzburg-Landau equations agrees well with the Landau level expansion 
and demonstrates that spatial distributions of the gap functions are different 
in the two bands and change with temperature. 
The temperature variation of the ratio of two gaps
is responsible for the upward temperature dependence of in-plane $H_{c2}$ 
as well as for the deviation of its out-of-plane behavior from the standard angular dependence. 
The hexagonal in-plane modulations of $H_{c2}$ can change sign with decreasing temperature.        
}

\maketitle

\section{Introduction}

Multigap superconductivity~\cite{suhl,moskalenko} has been discussed in the late 1950's 
for materials with a varying strength of electron-phonon interactions 
between different pieces of the Fermi surface. 
After the discovery of superconductivity in MgB$_2$~\cite{akimitsu} 
in 2001, an impressive 
collection of experimental and theoretical works~\cite{review} 
has established 
that this compound is the first unambiguous example of a
multigap superconductor. In MgB$_2$ 
the charge carriers are distributed between two sets of bands: the $\sigma$-bands with quasi-2D 
cylindrical Fermi sheets and the $\pi$-bands with 3D sheets forming a tubular network. 
The electron-phonon coupling is stronger in the $\sigma$-bands than in the $\pi$-bands, 
and gives rise to an $s$-wave phonon-mediated superconductivity with two gaps $\Delta_1 \sim 7$ meV 
and $\Delta_2 \sim 2.5$ meV. Since the two sets have different characteristics (interaction with 
phonons, geometry of the Fermi sheets, impurity dependence etc.), an interplay between them results 
in deviations from the standard BCS theory. The most striking consequences of the two gaps are 
the unusual anisotropic features of MgB$_2$ under magnetic field, for example, inequality between 
the penetration depth and the upper critical field anisotropies, and their variations with 
temperature~\cite{angst,samuely,cubitt1,shi,rydh,golubov1,golubov2,miranovic,kogan,dahm,gurevich,kita}, and 
the 30$^\circ$-reorientation of the flux line lattice with increasing magnetic field applied along 
the $c$-axis~\cite{cubitt2,zhitomirsky}.  

The two-gap Ginzburg-Landau (GL) theory for MgB$_2$ 
developed in Ref.~\cite{zhitomirsky}
(see also the preceding works~\cite{tilley,geilikman}) 
is the exact limit of the microscopic theory in the vicinity of
the transition temperature. It can thus account for 
most of the observed properties in a clear and coherent way near 
$T_c$, while its simplicity compared to earlier studies is useful 
to understand the physics in this material. In the present paper 
we extend our previous analysis of the two-band effects~\cite{zhitomirsky} 
on angular and temperature dependence of the upper critical field $H_{c2}$. 
We minimize the GL functional using a variational procedure, which 
highlights separate spatial anisotropies of the gap in each band. 
This is an improvement compared to the earlier solutions where only 
one common distortion for both gaps is considered~\cite{golubov2,dahm}.  
This method is compared to a solution based on the Landau level expansion. 
We then estimate the temperature range of the GL regime.  
The present study covers the out-of-plane $H_{c2}$ anisotropy. 
By going beyond the ellipsoid Fermi sheet approximation of
Refs.~\cite{miranovic,dahm}, we also calculate in-plane modulation 
of the upper critical field arising from the hexagonal crystal symmetry.       

For a clean two-band BCS superconductor with two gaps $\Delta_1$ and $\Delta_2$,
the GL functional~\cite{zhitomirsky} has the form 
\begin{eqnarray}
&& F_{GL} = \int dx \Bigl[ \alpha_1 |\Delta_1|^2  + \alpha_2 |\Delta_2|^2
- \gamma (\Delta_1^*\Delta_2 + \Delta_2^*\Delta_1)  \nonumber \\
&& \mbox{}\!+ K_{1i}|\Pi_i\Delta_1|^2\!+
K_{2i} |\Pi_i\Delta_2|^2 + \textstyle\frac{1}{2}\beta_1|\Delta_1|^4\!+
\textstyle\frac{1}{2}\beta_2|\Delta_2|^4 \!\Bigr], \nonumber \\
&& \Pi_i = -i \partial_i + \frac{2\pi}{\Phi_0} A_i , \ \
\alpha_{1,2} = \frac{g_{2,1}}{G}- N_{1,2}\ln\frac{2\omega_D e^C}{\pi T}, \nonumber \\
&&\gamma = \frac{g_3}{G}, \ \ K_{ni} = \frac{7\zeta(3)N_n}{16\pi^2 T_c^2}\langle v_{Fni}^2\rangle, \ \ \beta_n = \frac{7\zeta(3)N_n}{8\pi^2 T_c^2} \label{GL} 
\end{eqnarray}
where repeating index $i$ implies a sum, $\Phi_0$ is the quantum flux, $\bf A$ the potential vector, $g_1$ and $g_2$ the intraband pairing coefficients ($n$=1,2 for the $\sigma$,$\pi$-band), $g_3$ the interband pairing coefficient, $G=g_1 g_2 - g_3^2$, $N_n$ the density of states at the Fermi level in the band $n$, $\omega_D$ the Debye frequency, C the Euler constant, and $\langle v_{Fni}^2\rangle $ the square of the Fermi velocity $i$-component averaged on the sheet $n$. It is then convenient to write
$\alpha_1 = - a_1 t$ with $a_1=N_1$, $t = \ln(T_1/T)$
and $T_1= (2\omega_D e^C/\pi) e^{-g_2/GN_1}$ for the first active
band and $\alpha_2 = \alpha_{20} - a_2t$ with $a_2=N_2$ for the passive band. 
In MgB$_2$ the active and passive bands correspond to the $\sigma$ and $\pi$ bands,
respectively. 

The crystal structure of MgB$_2$ is uniaxial, so the gradient coefficients are the same for all directions in the basal plane, and $K_{na}=K_{nb}=K_{n}$. LDA calculations~\cite{kong} yield for 
the highly anisotropic $\sigma$-band $\langle v_{F1ab}^2\rangle = 2.13$ and 
$\langle v_{F1c}^2\rangle = 0.05$, while for the $\pi$-band $\langle v_{F2ab}^2\rangle = 1.51$ 
and $\langle v_{F2c}^2\rangle = 2.96$, all numbers are in units of $10^{15}$~cm$^2$/s$^2$. 
With the provided ratio $N_2/N_1=1.5$, the in-plane gradient constants for the two bands are 
practically the same $K_{2}/K_{1}\approx 1.06$, whereas the $c$-axis constants differ by almost 
two orders of magnitude $K_{2c}/K_{1c} \approx 90$.  A crude estimate for $H_{c2}$ 
at zero temperature by $\frac{N_1}{\sqrt{K_1 K_{1c}}} \frac{\Phi_0}{2\pi} \approx 4$T 
is substantially smaller than the experimental value $H_{c2}^{ab}(0 {\rm K}) \approx 18$T, 
which suggests the gradient constants based on LDA data are over-estimated by a factor of four. 
Such a discrepancy is due to a significant renormalization 
of effective masses by the electron-phonon coupling. 
The electron-phonon coupling leads to effective masses twice larger than 
the LDA prediction in the $\sigma$-band, 
whereas they are only slightly renormalized in the 
$\pi$-band~\cite{mazin}. 
The reduction of gradient term coefficients is given by
squares of the mass renormalization factors.

The interband impurity scattering in MgB$_2$ is exceptionally 
small due to its particular electronic structure, even in low 
quality samples \cite{mazin2}. The clean 
limit two-gap GL theory described above is straightforwardly extended 
to include the effect of s-wave intraband scattering by non-magnetic 
impurities~\cite{pokrovsky}: the GL functional keeps the same form 
wherein the expression for $K_{ni}$ has to be replaced 
by $K_{ni}= \pi N_n \langle v_{Fni}^2\rangle \Lambda(\tau_n)/8 T_c$ with  
\begin{equation}
\Lambda(\tau_n) = \tau_n \frac{8}{\pi^2} \sum_m \frac{1}{(2m+1)^2
((2m+1)2\pi \tau_n T_c +1)} \ , 
\end{equation}
where $\tau_n$ is the transport collision time in the band $n$. 
The intraband anisotropy is then the same as in the clean limit, 
while the renormalization factor $\Lambda(\tau_n)$ can vary between 
the two bands due to different sensitivity to impurities. 
The resulting GL equations are naturally found as the limit of 
Usadel equations near $T_c$.~\cite{golubov2,gurevich}.

\section{Upward curvature of $H_{c2}^{ab}(T)$}

In this section the $z$-axis is fixed along the crystal $c$-axis and the $y$-axis is taken parallel 
to the magnetic field applied in the $ab$-plane. The vector potential is chosen in the Landau gauge
as $ {\bf A} = (Hz,0,0)$. The coupled linearized GL equations for solutions homogeneous
along the field direction are
\begin{equation}
\left( \alpha_n + K_n h^2 z^2 - K_{nc} \partial_z^2 \right) \Delta_n - \gamma \Delta_{n'} = 0
\end{equation}
for $n=1,2$, $n'=2,1$, with the reduced magnetic field $h=2\pi H/\Phi_0$. 
Since $K_{1c}/K_1 \neq K_{2c}/K_2$, an analytic solution can not be obtained by rescaling distances 
as in the single-gap case. We, therefore, search for an approximate solution of the form
\begin{equation}
\left(\begin{array}{c}
\Delta_1 \\
\Delta_2 \end{array}\right)=\left(\begin{array}{c} c \;  \tilde{\xi}_1^{-\frac{1}{2}} \Psi_0(z/\tilde{\xi}_1) \\ d \; \tilde{\xi}_2^{-\frac{1}{2}} \Psi_0(z/\tilde{\xi}_2) \end{array}\right) 
\end{equation}
where the Landau level wave functions are defined by
\begin{equation}
\Psi_p(z)= \frac{1}{\sqrt{p!}} \left[ \frac{1}{\sqrt{2}} (-\partial_z +z )\right]^p \frac{e^{-z^2/2}}{\pi^{1/4}}
\end{equation}
Different coherence lengths for each band are allowed with the parameterization $\tilde{\xi}_n^2=\mu_n/h$ where $\mu_n$ quantifies the distortion of the spatial distribution of the $n$-th component (in the single-gap case, $\mu$ is the stretching factor of the flux line lattice at the upper critical field and is independent from temperature). The following quadratic form in the GL functional is then found:
\begin{eqnarray}
F_2 & = & (\alpha_1 + h\tilde{K}_1)|c|^2 + (\alpha_2 + h\tilde{K}_2)|d|^2 \nonumber \\
& & - \tilde{\gamma}(c^* d+ d^* c) \label{F2} 
\end{eqnarray}
with $ \tilde{K}_n = \textstyle\frac{1}{2} (K_n \mu_n + K_{nc}/\mu_n)$ and 
$\tilde{\gamma}^2 = \gamma^2 \frac{2\sqrt{\mu_1 \mu_2}}{\mu_1 + \mu_2}$. 
At the transition field, the determinant in Eq.~(\ref{F2}) vanishes. This condition leads to 
\begin{equation}
\tilde{h}(\mu_1,\mu_2) = - \frac{\alpha_1}{2\tilde{K}_1} - \frac{\alpha_2}{2\tilde{K}_2}
+  \sqrt{\left(\frac{\alpha_1}{2\tilde{K}_1}-\frac{\alpha_2}{2\tilde{K}_2}\right)^2\!
+\frac{\tilde{\gamma}^2}{\tilde{K}_1\tilde{K}_2}}
\label{eq:h-tilde}
\end{equation}
In order to find the (nucleation) upper critical field, $\tilde{h}$ is maximized
$\tilde{h}_{c2}=\max_{\mu_1,\mu_2} \; \tilde{h}(\mu_1,\mu_2)$.

\begin{figure}[t]
  \begin{center}
	  \includegraphics[width=0.95\columnwidth]{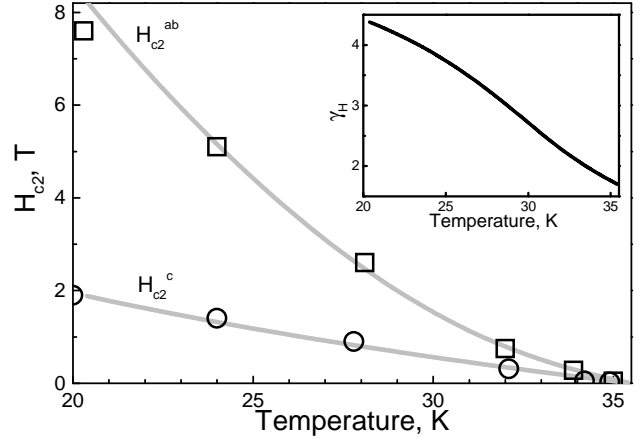}
	\end{center}  
	\caption{The upper critical field in MgB$_2$: experimental data from 
Lyard \emph{et al.}~\cite{samuely} (symbols) and the GL computations with parameters  $\alpha_{20}=0.65 a_1$ and $\gamma=0.4 a_1$ (solid lines). The inset shows the anisotropy ratio $\gamma_H=H_{c2}^{ab}/H_{c2}^c$.}
	\label{fig:040813}
\end{figure}

\begin{figure}[t]
\begin{center} 
\includegraphics[width=0.95\columnwidth]{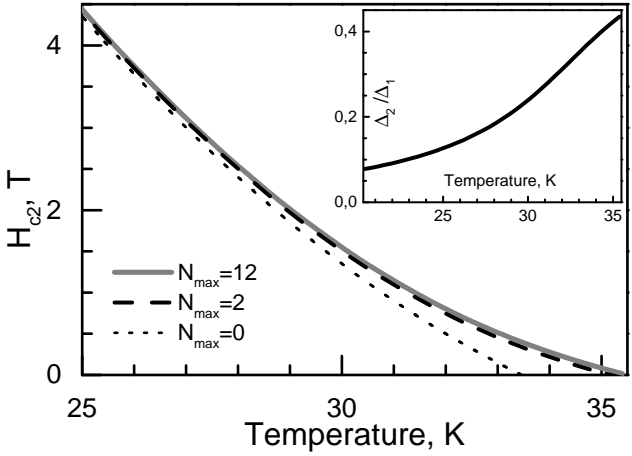}
\end{center}
\caption{In-plane $H_{c2}$ calculated with the Landau level expansion to the order $N_{\rm max}$ 
for the same parameters as in Fig.~\ref{fig:040813}. 
The inset displays the gap ratio $\rho=\Delta_2/\Delta_1$ found with the highest expansion order.}
  \label{fig:040815}
\end{figure}
Within the above variational scheme, the analytic expressions for the in-plane transition field 
are possible in two temperature regimes. Near $T_c$, vanishing $\tilde{h}$ implies that 
the superconducting gaps have the same variation length in each band.  
The condition $\mu_1=\mu_2$ yields
\begin{equation}
\tilde{h}^{ab}_{c2} \approx \frac{a_1 (t - t_c)}{\sqrt{(K_1+\rho^2 K_2)((K_{1z}+\rho^2 K_{2z})}}
\end{equation}
with the gap ratio $\rho=|d/c| \approx \sqrt{\alpha_1/\alpha_2}$ and $t_c=\ln (T_1/T_c)$. 
Since $K_{1c} \sim 0.01 K_{2c}$ in MgB$_2$, whereas $\rho^2 \approx 0.1$, we can simplify
the above expression to 
\begin{equation}
\tilde{h}^{ab}_{c2} \approx \frac{a_1 (t - t_c)}{\rho \sqrt{K_1 K_{2c}}} \; .
\end{equation} 
In the second temperature regime for $T<T_1$, the first active band is
dominant and 
\begin{equation}
\tilde{h}^{ab}_{c2} \approx \frac{a_1 t}{\sqrt{K_1 K_{1c}}} \; .
\end{equation} 
The line $H^{ab}_{c2}(T)$ exhibits, therefore, a marked upturn curvature between the two regimes, 
in contrast to $H_{c2}^c$(T). The two upper critical fields are plotted in Fig.~\ref{fig:040813}. 
In order to fit the experimental data,
we have renormalized all gradient constants 
obtained from the LDA data
by a factor of five.
The corresponding mass enhancement $\sqrt{5} \approx 2.2$ roughly agrees with 
the electron-phonon renormalization factor\cite{mazin}.
For simplicity, the same value has been applied for both bands. 

In order to verify an accuracy of the variational method, 
we alternatively proceed by expanding the gap functions in terms of the  
the Landau levels: $\Delta_n = \sum_p c_{n,p} \phi_{n,p}$ 
where $\phi_{n,p}(z)= \xi_n^{-1/2} \Psi_p(z/\xi_n)$ and $\xi_n^2= \sqrt{K_{nc}/K_{n}}/h$. 
For the upper critical field this expansion is restricted to the even order levels. 
The quadratic part of the GL functional has the following matrix element in this base:
\begin{eqnarray}
 && M^{F_2}_{2p+n,2q+n'} = \left( \alpha_n+(4p+1)h\sqrt{K_n K_{nc}} \right) 
\delta_{n,n'}\delta_{p,q}  \nonumber \\
& & -  \gamma \int \!\! dx \; \phi_{n,2p}^*(x) \phi_{n',2q}(x) \; (1-\delta_{n,n'})
\end{eqnarray}
with $n,n' \in \{ 1,2 \}$, and $p,q\geq0$. 
The upper critical field $h_{c2}$ is then approximated by the largest root of the sub-matrix 
determinant corresponding to the desired expansion up to the order $N_{\rm max}$.

Although the zeroth order approximation significantly
deviates near $T_c$ (see Fig.~\ref{fig:040815}), 
the procedure is rapidly converging with increasing 
the expansion order, even in the case of a 
great disparity between the two bands (e.g., $\xi_1^2/\xi_2^2 \geq 100$ 
or $\leq 0.01$). 
The expansion to the order $N_{\rm max}\geq 12$ yields the upper 
critical field curve
in excellent agreement 
with the variational solution (the two curves are indistinguishable
on the scale of Fig.~\ref{fig:040815}). 

\begin{figure}[t]
\begin{center}
\vspace{2 mm}
\includegraphics[width=0.95\columnwidth]{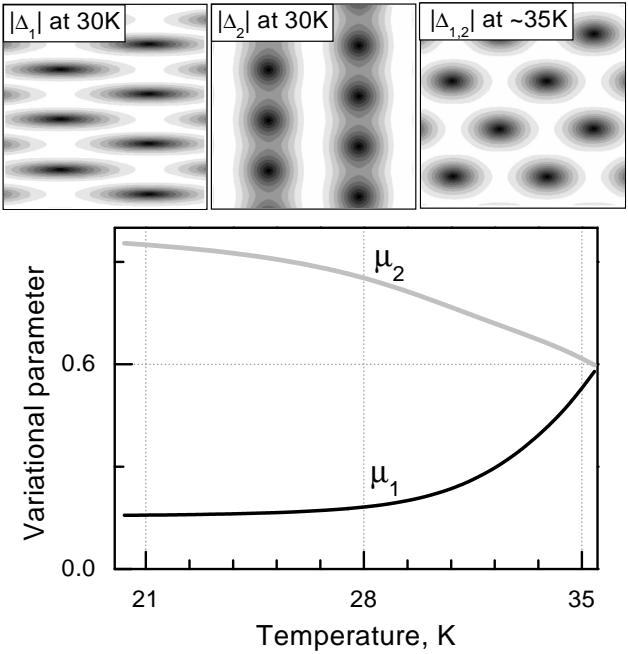}
\end{center}  
\caption{Lower panel: temperature dependence of 
variational parameters for in-plane magnetic field and 
the same set of GL parameters as in Fig.~\ref{fig:040813};
Upper panel: absolute values of the two gaps
in the vicinity of $H_{c2}$ near to and away from the transition
temperature.
}
\label{fig:040620-01b}
\end{figure}

Fig.~\ref{fig:040620-01b} displays the behavior of the parameters $\mu_n$ 
defining the effective anisotropy of the variation lengths 
$\tilde{\xi}_n$ in the plane perpendicular to the magnetic field, i. e. 
$\mu_n=\tilde{\xi}_{nc}/\tilde{\xi}_{nab}$ for the magnetic field applied 
in the basal plane. This confirms the above analytic predictions: 
the order parameter varies on different length scales for each band, 
and $\mu_n$ can change with temperature contrary to the single-gap case. 
At $T_c$, the two parameters have the same value 
$\sqrt{(K_{1z}+\rho^2 K_{2z})/(K_1 + \rho^2 K_2)}=0.59$ with $ \rho=0.44$, 
while $\mu_1 \approx\sqrt{K_{1c}/K_1}=0.15$ below $T_1=29$~K. 
We should stress that periodic vortex structures for the two gaps 
have the same lattice parameters for arbitrary ratio of $\mu_1/\mu_2$.
However, spatial distributions of $|\Delta_1(r)|^2$ and 
$|\Delta_2(r)|^2$  become quite different at low temperatures
once $\mu_1\ll \mu_2$.
Such a behavior is demonstrated on the top panel of 
Fig.~\ref{fig:040620-01b}. The different spatial distributions
of the two gaps can be probed by scanning tunneling microscopy.
Also, magnetic field generated by superconducting
currents $h_s(r) \sim (|\Delta_1(r)|^2+|\Delta_2(r)|^2)$
should deviate significantly for a distribution expected
for an anisotropic single-gap superconductor.
Muon spin relaxation measurements can in principle verify
such a behavior.

We shall now estimate the temperature range of the GL regime from the above computations. 
The gradient expansion is valid as long as $|K_{ni} \partial_i^2 \Delta_n| < \Delta_n$ 
for all $n$ and $i$. This condition is approximately replaced with $K_{ni}/\tilde{\xi}^2_{ni}<1$.
The most restrictive case is for $K_{2c}/\tilde{\xi}^2_{2c}=K_{2c} h_{c2}^{ab}/\mu_2$,
which becomes $\sim 1$ below $\sim 30$~K, well beyond 
a narrow temperature regime suggested for the GL theory by Golubov and Koshelev~\cite{golubov2}.
The discrepancy is partially terminological, since
in Ref.~\cite{golubov2}  the GL approximation always corresponds to
an effective (anisotropic) single-gap GL theory, which is correct only 
when the ratio of the two gaps is constant. 
As we have demonstrated above, the full two-gap GL theory is valid in a much wider temperature
range and describes adequately 
temperature variation of $\Delta_2/\Delta_1$ (Fig.~\ref{fig:040815}) and of 
the two coherence  lengths (Fig.~\ref{fig:040620-01b}).

\section{Angular dependence of out-of-plane $H_{c2}$}
 
\begin{figure}[t]
  \begin{center}
    \vspace{10 mm}
    \includegraphics[width=0.95\columnwidth]{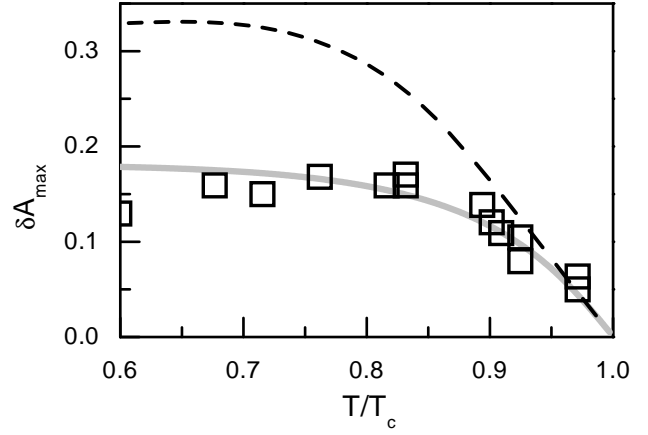}
  \end{center}  
	\caption{Maximum deviation from the single-gap GL scaling law : squares are experimental data~\cite{rydh} and lines are obtained from the two-gap GL theory.}
	\label{fig:040711}
\end{figure}

Let us now discuss the out-of-plane behavior of the upper critical field. 
In the single-gap anisotropic GL theory, when ${\bf H}$ is tilted from 
the c-axis by an angle $\theta$, 
the upper critical field has an elliptic (effective mass) angular 
dependence 
\begin{equation}
H_{\rm SAGL}(\theta,T) = 
\frac{H_{c2}^c(T)}{\sqrt{\cos^2(\theta)+\sin^2(\theta)\gamma_H^{-2}}} \ ,
\label{eq:sagl}
\end{equation}
where $\gamma_H = H^{ab}_{c2}/H^c_{c2}$ is a temperature independent 
constant
$\sqrt{K_{c}/K_{ab}}$. Experimental measurements in MgB$_2$ have shown 
that not only $\gamma_H$ 
changes with temperature (Fig.~\ref{fig:040813}) but deviations from 
the elliptic
angular dependence~(\ref{eq:sagl}) grow with decreasing 
temperature~\cite{angst,shi,rydh}. 
Such a behavior has been reproduced within quasi-classical Usadel equations~\cite{golubov2}. 
The methods we have employed for $H_{c2}^{ab}$ are still valid to find $H_{c2}(\theta)$: 
one needs only to replace $K_{nc}$ by an angular dependent 
$K_n(\theta) = \cos^2(\theta) K_{n} + \sin^2(\theta) K_{nc}$ in the previous formula. 
Expression~(\ref{eq:h-tilde}) for $\tilde{h}$ shows that the deviation grows with the disparity 
between the $\tilde{K}_n(\theta)$, so it increases when departing from $T_c$.
The deviations can be quantified by $\delta A (\theta)=1-(H_{c2}(\theta)/H_{\rm SAGL}(\theta))^2$. 
Fig.~\ref{fig:040711} displays the maximum deviation $\delta A_{\rm max} = 
{\rm max}_{\theta} \delta A (\theta)$. The dashed line is obtained from the two-gap GL theory with 
the parameters used above to fit the $H_{c2}$-data by Lyard \emph{et al.}\cite{samuely} 
in Fig.~\ref{fig:040813}. The calculation qualitatively reproduces experimental data from
Rydh \emph{et al.}~\cite{rydh}: $\delta A_{\rm max}$ increases with decreasing temperature 
and then saturates. But a quantitative discrepancy appears below $0.9T_c$ and becomes important 
at lower temperature. This deviation can be partially explained by the fact that experimental 
results are strongly sample dependent. At the present, the origin of the discrepancy remains 
an opened question. The full line is obtained with a modified interband 
coupling $\gamma^\star=0.3 a_1$ and $K_{c2}^\star=K_{2c}/3$ corresponding to a smaller anisotropy 
in the $\pi$-band.  

\section{In-plane modulation of $H_{c2}$}

In a hexagonal crystal, the transition magnetic field should exhibit a six-fold modulation 
when rotated about the $c$-axis~\cite{skokan}. 
The crystal field effect on superconductivity can be incorporated to the GL theory by 
including higher order (non-local) gradient terms~\cite{hohenberg}. 
Symmetry arguments suggest that coupling between the superconducting order parameter 
and the hexagonal crystal lattice appears at the sixth-order gradient terms. 
For a two-gap superconductor like MgB$_2$, the additional sixth-order part of the free energy 
is a sum of separate contributions from each band: $F_{GL} \longrightarrow 
F_{GL} + F_{6,1} + F_{6,2}$. The correction derived from the BCS theory~\cite{zhitomirsky,takanaka} 
is (omitting the index $n=1,2$ for brevity)
\begin{eqnarray}
F_6 & = &
\frac{\zeta(7)N}{32\pi^6T_c^6}
\left(1-\frac{1}{2^7}\right)
\langle v_{Fi} v_{Fj}v_{Fk}v_{Fl}v_{Fm}v_{Fn}\rangle
\nonumber \\
& & \mbox{} \times
(\Pi_i\Pi_j\Pi_k\Delta)^*(\Pi_l\Pi_m\Pi_n\Delta) \ .
\end{eqnarray}
Setting the $z$-axis perpendicular to the basal plane, the above terms can be split into 
isotropic in-plane part 
\begin{equation}
F_6^{\rm iso} =  K_6^{i}\Delta^*\Bigl[\Pi_x^2+\Pi_y^2\Bigr]^3 \Delta
\end{equation}
with $K_6^{i} = \frac{\zeta(7)N}{64\pi^6T_c^6} \left(1-\frac{1}{2^7}\right)
\bigl(\langle v_{Fx}^6\rangle + \langle v_{Fy}^6\rangle\bigr)$, 
and anisotropic in-plane contribution
\begin{equation}
F_6^{\rm an} = \frac{1}{2}K_6^{a}\Delta^*\Bigl[(\Pi_x+i\Pi_y)^6+(\Pi_x-i\Pi_y)^6
\Bigr]\Delta 
\end{equation}
with $K_6^{a} = \frac{\zeta(7)N}{64\pi^6T_c^6} \left(1-\frac{1}{2^7}\right) \bigl(\langle v_{Fx}^6\rangle - \langle v_{Fy}^6\rangle\bigr)$.
This expression of $F^{\rm an}_6$ assumes that the $x$- and the $y$-axes are parallel to
the reflection lines in the $ab$-plane. With the $x$-axis parallel to the $b$-direction, 
tight-biding calculations\cite{zhitomirsky}
 yield $\langle v_{Fx}^6 \rangle = 4.608$, $\langle v_{Fy}^6 \rangle 
= 4.601$ for the $\sigma$-band, while for the $\pi$-band, $\langle v_{Fx}^6 \rangle = 1.514$, 
$\langle v_{Fy}^6 \rangle = 1.776$ in units of $10^{46}$ (cm/s)$^6$. 
The different sign of the hexagonal harmonics of the Fermi velocities in the two bands
is responsible for a unique $30$-degree orientational transition of the vortex
lattice in MgB$_2$.\cite{zhitomirsky}
No theory can describe at present the electron-phonon effect on the hexagonal modulation 
of the Fermi surface. We use, therefore, the raw LDA values for all gradient coefficients 
in the consideration below. If we rotate now the orthogonal axes so that the $y$-axis is parallel to the magnetic field ${\bf H}$ when the latter forms an angle $\phi$ with the $a$-axis, the terms in $F_6$ change in a simple way: $F_6^{\rm iso}$ is preserved while $F_6^{\rm an}$ turns into 
\begin{equation}
F_6^{\rm an}= \frac{1}{2} K_6^{a}\Delta^*\Bigl[e^{i6\phi}(\Pi_x+i\Pi_y)^6+e^{-i6\phi}(\Pi_x-i\Pi_y)^6
\Bigr]\Delta \
\end{equation}

\begin{figure}[t]
  \begin{center}
	  \includegraphics[width=0.95\columnwidth]{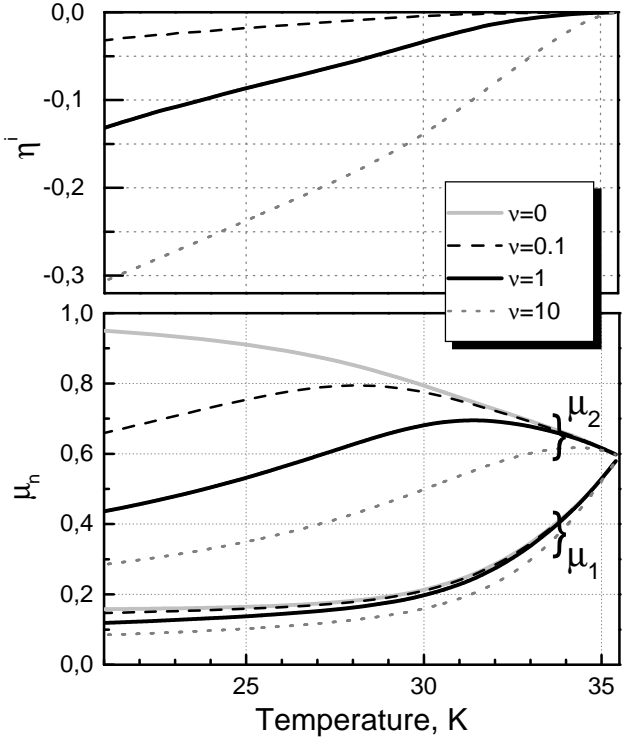}
	\end{center}  
	\caption{Influence of the isotropic correction $F_6$ with $(\langle v_{F1}^6 \rangle,\langle v_{F2}^6 \rangle)=\nu (4.6,1.6)$ (in units of $10^{46}$ (cm/s)$^6$) for different magnitudes $\nu$: relative shift of the upper critical field $\eta^i=(H_{c2}^{\nu}-H_{c2}^{\circ})/H_{c2}^{\circ}$ and corresponding variational parameters $\mu_n$.}
	\label{fig:040622-01}
\end{figure}
Since $\Pi_y=0$, the extra term can be written as 
$F_6= \sum K_{6,n}\Delta_n^* \Pi_x^6 \Delta_n$ with $K_{6,n}=K_{6,n}^{i}+K_{6,n}^{a}\cos (6\phi)$.
For the variational approximation, the new functional yields the  quadratic form 
\begin{eqnarray}
F_2 & = & (\alpha_1 +  \tilde{K}_1 h + \tilde{K}_{6,1} h^3) |c|^2 + (\alpha_2 + \tilde{K}_2 h  \nonumber \\
& & + \tilde{K}_{6,2} h^3) |d|^2 - \tilde{\gamma}(c^* d+ d^* c) 
\end{eqnarray}
with $ \tilde{K}_{6,n} = \frac{15}{8} K_{6,n} \mu_n^3$. While in the expansion method, this results in the new matrix element 
\begin{equation}
 M^{F_2+F_6}_{2p+n,2q+n'} =  M^{F_2}_{2p+n,2q+n'} + h^3 K_{6,n} \left( \frac{K_{nc}}{K_n} \right)^{\!\! 3/2} \!\!\!\! M^{(6)}_{2p,2q} \delta_{n,n'} 
\end{equation}
with $M^{(6)}_{p,q}= \frac{1}{8}\left\langle \Psi_p|( \hat{a}^{\dagger} + \hat{a} )^6|\Psi_q \right\rangle$ where $\hat{a}$ is the annihilation operator of Landau levels.

In the weakly anisotropic regime $F_6 <\!\!< F_2$, we expect
\begin{equation}
H_{c2}(\phi)\approx H_{c2}^{\circ} \left( 1+\eta^i+\eta^a \cos (6\phi)\right)
\end{equation}
The isotropic parts yield a $\phi$-independent shift of $H_{c2}$ (and ensure $K_{6,n} >0$ for the numerical solution converging) while the anisotropic parts are responsible for the six-fold modulation of the correction. $\eta^a$ can change sign when the temperature varies because the anisotropies in each band are opposite. Fig.~\ref{fig:040622-01} displays the corrections brought by the isotropic parts of $F_6$. The deviations become important below 30K as expected out of the estimated GL regime, which implies the necessity to retain higher order terms in the gradient expansion of the GL functional.       

The extra $h^3$ terms prevent from deriving an analytical expression for the magnetic field correction $\delta h_{c2}=h_{c2}(\phi)-h_{c2}(\pi/12)$. We can however partially estimate the latter. Let us name the quantities related to the quadratic form $(F_{2}+F_6^{\rm iso})$ with the superscript "$\circ$", and the ones for $(F_2+F_6^{\rm iso}+F_6^{\rm an})$ without it. Within the variational method, we then find with a perturbation expansion 
\begin{eqnarray}
\frac{\delta \tilde{h}_{c2}}{\tilde{h}^{\circ}_{c2}} & \approx & \frac{1}{\tilde{h}^{\circ}_{c2}} \frac{d \tilde{h}^{\circ}_{c2}}{dT} \left. \frac{\langle \Delta^{\circ} | F_6^{\rm an} | \Delta^{\circ} \rangle}{\langle \Delta^{\circ} | \partial (F_2+F_6^{\rm iso})/\partial T |\Delta^{\circ} \rangle} \right|_{\tilde{h}_{c2}^{\circ}} \label{eq:estimation1} \\
& \approx & \frac{15}{8}\cos (6\phi) T\frac{d \tilde{h}^{\circ}_{c2}}{d T}\tilde{h}_{c2}^{{\circ} \; \stackrel{\scriptstyle 2}{\;} } \frac{ K_{6,1}^a \mu_{1}^{{\circ} \; \stackrel{\scriptstyle 3}{\;}} + K_{6,2}^a \mu_{2}^{{\circ} \; \stackrel{\scriptstyle 3}{\;}} \rho^2}{ a_1 + a_2 \rho^2}  \nonumber 
\end{eqnarray}
where $\rho^2= |c^{\circ}/d^{\circ}|^2$. The expansion method provides in a similar way
\begin{equation}
\frac{\delta h_{c2}}{h^{\circ}_{c2}} \approx \frac{T}{h^{\circ}_{c2}} \frac{d h^{\circ}_{c2}}{dT} \left. \frac{(1+\rho^2)\langle \Delta^{\circ} | F_6^{\rm an} | \Delta^{\circ} \rangle}{a_1 + a_2 \rho^2} \right|_{h_{c2}^{\circ}}
\label{eq:estimation2}
\end{equation}
but $\langle \Delta^{\circ} | F_6^{\rm an} | \Delta^{\circ} \rangle$ has a more complicated expression.

\begin{figure}[t]
  \begin{center}
	  \includegraphics[width=0.95\columnwidth]{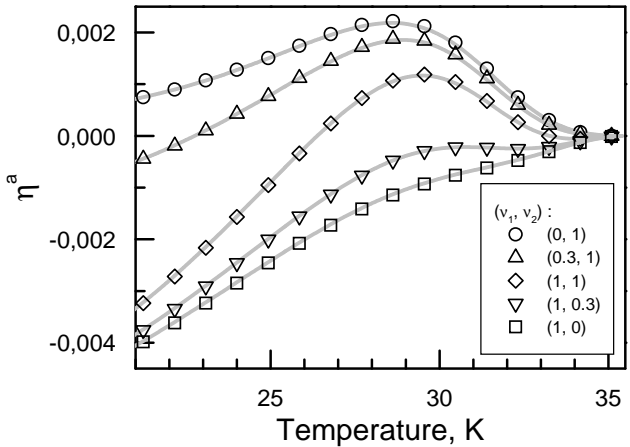}
	\end{center}
	\caption{Relative modulation amplitude for different pairs $(\nu_1,\nu_2)$: solid lines are obtained with the variational method and symbols with estimate (\ref{eq:estimation1}).}
	\label{fig:040622-02-2}
\end{figure}

In Fig.~\ref{fig:040622-02-2}, we have plotted the relative modulation amplitude $\eta^a=(h_{c2}(0)-h_{c2}(\pi/6))/(h_{c2}(0)+h_{c2}(\pi/6))$ with the hexagonal anisotropy $\langle v_{Fb}^6 \rangle = \langle v_{F}^6 \rangle^i + \langle v_{F}^6 \rangle^a$ and $\langle v_{Fa}^6 \rangle = \langle v_{F}^6 \rangle^i - \langle v_{F}^6 \rangle^a$ where $(\langle v_{F1}^6 \rangle^i,\langle v_{F2}^6 \rangle^i)=(4.6,1.6)$ while $(\langle v_{F1}^6 \rangle^a,\langle v_{F2}^6 \rangle^a)=(0.3\nu_1,-0.2\nu_2)$ (in units of $10^{46}$ (cm/s)$^6$). Ab initio calculations provides $\langle v_{F1}^6 \rangle^a \ll \langle v_{F2}^6 \rangle^a$ for MgB$_2$ which corresponds around to the couple $(\nu_1,\nu_2)=(0;1)$ in Fig.~\ref{fig:040622-02-2}. Due to the LDA results uncertainty and also to illustrate the interplay between the two bands, the plots for other values of $(\nu_1,\nu_2)$ are displayed. Note the results at low temperature should be taken with caution since they are obtained out of the GL regime. When the hexagonal anisotropies of each band are of the same order, $\eta^a$ sign can change with temperature. But this modulation is too small to be detected experimentally in the GL regime, which agrees with measurements reported by Shi \emph{et al.}~\cite{shi}. Estimation~(\ref{eq:estimation1}) gives three reasons for this. First, $\eta^a$ grows as $h_{c2}^2$ contrary to the four-fold symmetry crystal case where the increase is linear. Then the anisotropies of the two bands oppose each other. And finally, even though $\langle v_{F1}^6 \rangle^a$ would be too small to compete with $\langle v_{F2}^6 \rangle^a$, the contribution from the second band is reduced by the rapidly decreasing factor $\rho^2$ and, below 30K, by $\mu_2^3$.   
 
\section{Conclusions}

Angular and temperature dependence of the upper critical field of MgB$_2$
have been determined within the two-gap GL theory. We have used two different numerical methods 
which are in excellent agreement with each other and yield an unconventional anisotropy of $H_{c2}$ 
observed in the superconductor MgB$_2$. Such a behavior reflects the different Fermi sheet 
geometries and the varying importance of the small $\pi$-gap. The zeroth Landau levels 
employed in the variational approach are sufficient for accurate description
of the continuous transition at $H_{c2}$.
Contrary to the single-gap case, spatial anisotropy of the  gap functions
in the plane perpendicular to the magnetic field changes with temperature and 
can be different for each band. 
This explains the deviation from the effective mass angular dependence~(\ref{eq:sagl})
applicable to ordinary superconductors. Existence of two different 
characteristic lengths should also affect 
the vortex core shape,\cite{zhitomirsky}
especially when an applied field is perpendicular to the $c$-axis. 
The gap functions have an effective single-component behavior only in a temperature region 
near $T_c$ significantly narrower than the range for the validity of the two-gap GL theory 
$\sim (T_c-T)/T_c \sim 1/7$. At last, the hexagonal $ab$-plane modulation 
of $H_{c2}$ arising from the crystal symmetry 
can result in a change of 
the sign of the hexagonal harmonics of $H_{c2}(\theta)$ when the temperature is decreased.

\end{document}